\documentclass[12pt]{article}
\usepackage{amsmath}
\usepackage{amssymb}
\usepackage[dvips]{graphicx}
\usepackage{amsmath,amscd}
\usepackage{color}
\usepackage{mathrsfs}
\usepackage{latexsym}
\usepackage{ulem}
\makeatletter

  \@addtoreset{equation}{section}
\makeatother
\usepackage[dvips]{graphicx}

\begin{document}

\begin{center}
{\LARGE Matter scattering in $R_{\mu \nu}^2$ gravity and unitarity}
\end{center}
\begin{center}
\lineskip .45em
\vskip1.5cm
{\large Yugo Abe$^a$\footnote{E-mail: yugoabe@miyakonojo.kosen-ac.jp}, 
Takeo Inami$^{b}$\footnote{E-mail: inami@phys.chuo-u.ac.jp},
Keisuke Izumi$^{c}$\footnote{E-mail: izumi@math.nagoya-u.ac.jp},
and Tomotaka Kitamura$^{\rm d}$\footnote{E-mail: kitamura@gravity.phys.waseda.ac.jp}}

\vskip 1.5em
${}^a\,${\itshape National Institute of Technology, Miyakonojo College, Miyakonojo 885-8567, Japan}\\[1mm]
${}^b\,${\itshape Theoretical Research Division, Nishina Center, RIKEN, Wako 351-0198, Japan}\\[1mm]
${}^b\,${\itshape Department of Physics, Sungkyunkwan University, Suwon 16419, Republic of Korea}\\[1mm]
${}^c\,${\itshape Kobayashi-Maskawa Institute, Nagoya University, Nagoya 464-8602, Japan}\\[1mm]
${}^c\,${\itshape Department of Mathematics, Nagoya University, Nagoya 464-8602, Japan}\\[1mm]
${}^d\,${\itshape Department of physics, Waseda University, Shinjyuku, Tokyo 169-8555, Japan}
\end{center}
\begin{abstract}
We investigate the ultraviolet (UV) behavior of two-scalar elastic scattering with graviton exchanges in higher-curvature gravity theory.
In Einstein gravity, matter scattering is shown not to satisfy the unitarity bound at tree level at high energy.
Among some of the possible directions for the UV completion of Einstein gravity, such as string theory, modiﬁed gravity,
and inclusion of high-mass/high-spin states, we take $R_{\mu\nu}^2$ gravity coupled to matter.
We show that matter scattering with graviton interactions satisﬁes the unitarity bound at high energy,
even with negative norm states due to the higher-order derivatives of metric components.
The difference in the unitarity property of these two gravity theories is probably connected to that in another UV property, namely, the renormalizability property of the two.
\end{abstract}
\newpage
\tableofcontents

\section{Introduction}
\label{sec:intro}
The fact that Einstein gravity is not ultraviolet (UV) renormalizable due to 
the negative mass dimension of the coupling constant $\kappa$ prevents us from evaluating quantum gravity effects in a reliable manner. 
It is commonly believed that if a quantum field theory (QFT), including gravity theory,
is renormalizable, scattering amplitudes of this theory satisfy unitary at high energy, and vice versa.
This property has been studied in a few QFTs, particularly in (massive) gauge theories \cite{Bell,L.Smith,CLT}.
The same property has scarcely been studied in quantum gravity theory, for the reason that UV complete theories of gravity are poorly understood. 
Recently, the UV completion of the gravity from the viewpoint of unitarity on gravitational scattering has been discussed \cite{Yu-tin}. 

In this letter, we undertake a study of the high-energy unitarity property in quantum gravity theories with supposedly good UV behaviors.
In particular, we study the unitarity bound, which is an inequality related to scattering amplitudes and the necessary condition for unitarity \cite{CLT,FIIK1}, in $R^2$ gravity.
Here $R^2$ gravity means the theory with $R_{\mu \nu}^2$ as well as $R^2$, also called higher-curvature gravity. 
$R^2$ gravity is shown to be renormalizable \cite{Stelle}, but the existence of matter fields would make the discussion complicated.

The theory has negative norm states because of the higher-order derivatives of the metric components.
Although complete unitarity is not satisﬁed due to the negative norm states,
we suspect that there is a relation between the renormalizability and the UV behavior of scattering amplitudes, i.e., the unitarity bound.
We focus on the unitarity bound as a necessary condition for unitarity at high energies,
and we study the elastic scattering amplitude in theories with gravitational interactions.

In the study of gauge theories of strong interactions and of electroweak interactions,
matter scattering plays a vital role in unveiling the asymptotically free nature of the former theory and the renormalization property of the latter.
We mean deep inelastic $e$--$p$ scattering in the former \cite{Feynman} and neutrino scattering in the latter \cite{AL}. 
In view of these experiences, we begin by studying matter scattering amplitudes in $R^2$ gravity
and compare them with the non-unitarity property of the scattering amplitudes in Einstein gravity.
We will see that the matter scattering amplitudes satisfy the unitarity bound, even though the theory has negative norm states,
suggesting that the addition of the $R^2$ term is one possibility for UV completion.
\section{Summary of the matter sector coupled to Einstein gravity}
\label{sec:Einstein gravity}
\begin{figure}[h!]
\begin{center}
\includegraphics[width=35mm]{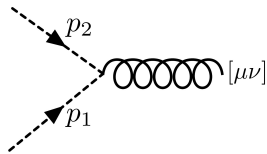}
\vspace{0mm}
\vskip-\lastskip
\caption{The matter--graviton three-point ($h_{\mu\nu}\phi\phi$) vertex function $\lambda_{\mu\nu}$.}
\label{fig:3-point vertex}
\end{center}
\end{figure}

We compute the matter scattering amplitude due to graviton exchanges.
We consider the $\phi ^4$ theory coupled to Einstein gravity on a 4D Minkowski background.
The action is written as    
\begin{align}
S=\int d^{4}x\sqrt{-g}\left(-\frac{1}{\kappa^{2}}R+\frac{1}{2}\nabla_{\mu}\phi\nabla^{\mu}\phi-\frac{1}{2}m^{2}\phi^{2}-\frac{1}{4!}\lambda\phi^{4}\right), 
\end{align}
where $R$ is the curvature scalar, $\kappa^{2}$ is the coupling constant with dim $[\kappa^{2}]=-2$, and $m$ is the mass of $\phi$.
We use the $(+,-,-,-)$ convention for the metric. 
The gravitational field is defined as the fluctuation of the physical metric,
$h_{\mu\nu}:=g_{\mu\nu}-\eta_{\mu\nu}$, from the background Minkowski metric $\eta_{\mu\nu}$. 
The Feynman rule for $h_{\mu\nu}$ can be found in Refs. \cite{DeWitt3,BG}.
In the calculation of the elastic $\phi$--$\phi$ scattering,
we need the graviton propagator and the $h_{\mu\nu}\phi\phi$ vertex.
For computational simplicity we take the propagator in the de Donder gauge:
\begin{align}
G^{\mu\nu\rho\sigma}=\frac{\kappa^{2}}{p^{2}}\left(\eta^{\mu\rho}\eta^{\nu\sigma}+\eta^{\mu\sigma}\eta^{\nu\rho}-\eta^{\mu\nu}\eta^{\rho\sigma}\right).
\label{Einstein propagator}
\end{align}
Note that scattering amplitudes are on-shell quantities and hence they should not depend on the gauge.
The $h_{\mu\nu}\phi\phi$ vertex is 
\begin{align}
\lambda_{\mu\nu}&:=\frac{-i}{2}\left(\eta_{\mu\nu}\left(p_{1_\alpha} p_{2}^\alpha+m^{2}\right)+p_{1\mu}p_{2\nu}+p_{2\mu}p_{1\nu}\right)\nonumber\\
&=\frac{-i}{4}\left(\theta_{\mu\nu}p^2+\left(p_{1\mu}-p_{2\mu}\right)\left(p_{1\nu}-p_{2\nu}\right)+\eta_{\mu\nu}(p_{1}^{2}+p_{2}^{2}-2m^{2})\right),
\label{3-point vertex}
\end{align}
where $p_{\mu}=p_{1\mu}+p_{2\mu}$, $p_{i\mu}$ $(i=1,2)$ are the 4-momenta of scalar fields (see Fig.\ref{fig:3-point vertex}),
and $\theta_{\mu\nu}$ is the projection tensor for vectors on the hypersurface normal to $n^{\mu}$, defined as 
\begin{align}
\theta_{\mu\nu}:=\eta_{\mu\nu}- \frac{p_\mu p_\nu}{p^2}.
\label{theta}
\end{align}
The last term in Eq.(\ref{3-point vertex}) becomes zero, if $p_{1\mu}$ and $p_{2\mu}$ satisfy the on-shell condition. 
\vspace{0mm}

\begin{figure}[h!]
\begin{center}
\includegraphics[width=100mm]{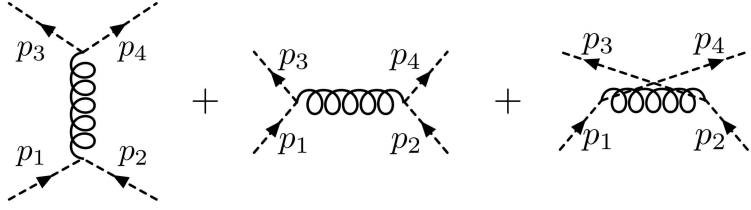}
\vspace{0mm}
\vskip-\lastskip
\caption{$s$-, $t$-, $u$-channel graviton exchanges in $\phi$--$\phi$ scattering.}
\label{fig:matter scattering}
\end{center}
\end{figure}

We consider the elastic $\phi$--$\phi$ scattering
\begin{align}
\phi+\phi\rightarrow\phi+\phi,
\label{the matter scattering}
\end{align}
due to graviton exchanges in the $s$-, $t$-, $u$-channels, as shown in Fig.\ref{fig:matter scattering}. 
\vspace{0mm}

We compute the amplitude $M_{s}(\phi\phi\rightarrow\phi\phi)$ due to the $s$-channel exchange
and express it in the Mandelstam variables $s$, $t$, and $u$:
\begin{align}
M_{s}=2\kappa^{2}\left[\frac{tu}{s}+2m^{2}-2\frac{m^{4}}{s}\right].
\label{Ms}
\end{align}
The other two terms $M_{t}$, $M_{u}$ due to the $t$- and $u$-channel exchanges are obtained from $M_{s}$ by the crossing relations
$s \leftrightarrow t$ and $s \leftrightarrow u$, respectively.
The sum of the three terms and the contact term from the direct $\phi^{4}$-coupling is 
\begin{align}
M(\phi\phi\rightarrow\phi\phi)=-\lambda+2\kappa^{2}\left[\left(\frac{tu}{s}+\frac{su}{t}+\frac{st}{u}\right)+6m^{2}-2m^{4}\left(\frac{1}{s}+\frac{1}{t}+\frac{1}{u}\right)\right].
\end{align}

We examine the unitarity bound of this amplitude in the high-energy limit:
\begin{align}
s\rightarrow\infty,~~{\rm with~\theta~being~fixed}.
\end{align}
Here $\theta$ is the scattering angle in the center-of-mass frame.
Recalling that
\begin{align}
t\simeq-\frac{1}{2}s(1-\cos{\theta}),~~~~~u\simeq-\frac{1}{2}s(1+\cos{\theta}),
\end{align}
in the high-energy limit,
we obtain 
\begin{equation}
\begin{split}
M&\simeq-\lambda+{\kappa}^{2}[2sf(\cos{\theta})+{\cal O}(s^{0})]\\
&={\cal O}(s^{1}). 
\end{split}
\label{s^1}
\end{equation}
Here we have set 
$f(x)=\frac{1}{4}(1-x^{2})+2(1+x^{2})/(1-x^{2})$.

If the theory is unitary, the elastic scattering amplitudes $M$ of two scalar fields are bounded by some constant $C$ at all energies;
namely, one can choose some constant $C$ such that $\left|M\right|<C$ holds at all energies \cite{FIIK1}.
The energy dependence in Eq.(\ref{s^1}), $M = {\cal O}(s^{1})$, means that the scattering amplitude exceeds the constant at some energy,
and thus the matter scattering amplitude does not obey high-energy unitarity.
It has been pointed out earlier that matter--graviton scattering amplitudes do not satisfy unitarity \cite{BG}.
Both results are in accord with the fact that Einstein gravity is not UV renormalizable.
\section{$R^{2}$ gravity coupled to a scalar field $\phi$}
\label{sec:R2 gravity}
Four decades ago Stelle proposed $R^2$ gravity as a renormalizable quantum gravity \cite{Stelle, VT}.
Perturbative renormalizability and the asymptotically free property have subsequently been studied \cite{FT, Tomboulis, JT}.
$R^2$ gravity with matter has been studied by a power-counting argument \cite{MO}.
The same question for the matter Lagrangian due to graviton one-loops has recently been studied \cite{SS,AHI,Horikoshi}.
In this situation of incomplete understanding of the UV properties of $R^{2}$ gravity,
an investigation of the high-energy unitarity of the scattering amplitudes in $R^{2}$ gravity should be very useful.
We will see that a study of the unitarity bound of the matter scattering amplitude already gives an insight into this question.
The higher-curvature terms also appear as an effective gravity action from the closed string theory \cite{GS,BR,HO,GKKLR}.
A connection to scattering amplitudes in strings may also be implied from this point of view. 

We recapitulate the $R^{2}$ gravity coupled to a scalar field $\phi$.
The action is given by 
\begin{align}
S&=\int d^{4}x\sqrt{-g}\left(-\frac{1}{\kappa^{2}}R+\alpha R^{2} +\beta R_{\mu \nu}^{2}+\frac{1}{2}\nabla_{\mu}\phi \nabla^{\mu}\phi -\frac{1}{2}m^{2}\phi^{2}-\frac{1}{4!}\lambda\phi^{4}\right),\\
&=\int d^{4}x\sqrt{-g}\left(-\frac{1}{\kappa^{2}}R+\frac{1}{3}(3\alpha+\beta)R^{2} +\frac{1}{2}\beta C_{\mu \nu\alpha\beta}^{2}-\frac{1}{2}\beta GB \right. \nonumber \\
&\qquad\qquad\qquad\qquad\qquad\qquad\qquad\qquad\left.+\frac{1}{2}\nabla_{\mu}\phi\nabla^{\mu}\phi -\frac{1}{2}m^{2}\phi^{2}-\frac{1}{4!}\lambda\phi^{4}\right),
\label{Wformaction}
\end{align}
where $\alpha$ and $\beta$ are dimensionless parameters of the theory, expected to be ${\cal O}(1)$.
$C_{\mu\nu\alpha\beta}$ is the Weyl tensor, i.e.,
\begin{align}
C_{\mu\nu\alpha\beta}:=R_{\mu\nu\alpha\beta}-\frac{1}{2}\left(g_{\mu\alpha}R_{\nu\beta}+g_{\nu\beta}R_{\mu\alpha}-g_{\mu\beta}R_{\nu\alpha}-g_{\nu\alpha}R_{\mu\beta}\right)+\frac{1}{6}R\left(g_{\mu\alpha}g_{\nu\beta}-g_{\mu\beta}g_{\nu\alpha}\right),
\end{align}
and $GB$ shows the Gauss--Bonnet term, i.e.,
\begin{align}
GB:=R_{\mu\nu\alpha\beta}^{2}-4R_{\mu\nu}^{2}+R^{2}.
\end{align}
The Gauss--Bonnet term is a topological term, and thus we can remove it from the action.
Because of the higher-order derivative terms of the metric components, the Feynman rules for graviton are quite involved.
For the purpose of computing the $\phi$--$\phi$ scattering amplitude, it suffices to know the $h_{\mu\nu}\phi\phi$ vertex and the graviton propagator.
The form of the $h_{\mu\nu}\phi\phi$ vertex is the same as that in the case of Einstein theory, Eq.(\ref{3-point vertex}), because the vertex stems 
from the matter parts of the action.%
\footnote{For multiplicative renormalizability, actually $\xi R\phi^{2}$ and the cosmological constant $\Lambda$ are also required \cite{EOR}.
However, we drop them, since it is possible to verify separately that each term satisfies the unitary bound.}

The graviton propagator is obtained in the usual manner,
\begin{equation}
\begin{split}
&G_{\mu\nu,\alpha\beta}=\frac {2}{\beta p^{4}+\kappa^{-2}p^{2}} P^{(2)}_{\mu\nu,\alpha\beta}+ \frac{1}{2(3\alpha+\beta)p^4-\kappa^{-2} p^2 }P^{(0)}_{\mu\nu,\alpha\beta},\\
&P^{(2)}_{\mu\nu,\alpha\beta}:= \frac{1}{2} \left(\theta_{\mu\alpha}\theta_{\beta\nu}+\theta_{\mu\beta}\theta_{\alpha\nu}\right) -\frac{1}{3}\theta_{\mu\nu}\theta_{\alpha\beta},\\
&P^{(0)}_{\mu\nu,\alpha\beta}:= \frac{1}{3}\theta_{\mu\nu}\theta_{\alpha\beta},
\label{R2 propagator}
\end{split}
\end{equation}
where the projection tensor $\theta_{\mu\nu}$ has already been defined in Eq.(\ref{theta}).
$P^{(2)}_{\mu\nu,\alpha\beta}$ and $P^{(0)}_{\mu\nu,\alpha\beta}$ are the projection tensors
for the symmetric rank-2 tensor to the spin-2 and (part of the) spin-0  components, respectively.
Here, we only show the gauge-independent parts of the propagator,
which are obtained by operating the projection tensor $\theta_{\mu\nu}$ on all indices of the propagator with any gauge fixing. 
In the calculation of the tree-level amplitude, the result depends only on the gauge-independent parts.%
\footnote{In the exact derivation of the propagator, we have to introduce the gauge-fixing term and ghost modes.
For the complete form, see Refs.\cite{BOS}.}

The propagator in Einstein gravity is regained by setting $\alpha=\beta=0$:
\begin{align}
G^{\rm E}_{\mu\nu,\alpha\beta}&=\frac{2\kappa^{2}}{p^{2}}P^{(2)}_{\mu\nu,\alpha\beta}-\frac{\kappa^{2}}{p^{2}}P^{(0)}_{\mu\nu,\alpha\beta}=\frac{\kappa^{2}}{p^{2}} \left(\theta_{\mu\alpha}\theta_{\beta\nu}+\theta_{\mu\beta}\theta_{\alpha\nu}-\theta_{\mu\nu}\theta_{\alpha\beta}\right), 
\label{Einstein limit}
\end{align}
which coincides with the projected propagator, i.e., replacing $\eta_{\mu\nu}$ by $\theta_{\mu\nu}$, of Eq.(\ref{Einstein propagator}).%
\footnote{The propagator in the  de Donder gauge contains gauge-dependent parts. Only the projected parts in the propagator are independent of the gauge parameters.}
We can see the difference of the UV behavior in the propagators, i.e., the propagator in Einstein gravity is ${\cal O}(p^{-2})$,
while that in $R^2$ gravity behaves as ${\cal O}(p^{-4})$ in the UV limit. 
Note that the coefficients of $p^4$ in the denominators for the spin-2 and spin-0 parts in the propagator (\ref{R2 propagator}) are proportional to those of $C_{\mu\nu\alpha\beta}^2$ and $R^2$ in the action (\ref{Wformaction}), respectively. 
Hence, the $C_{\mu\nu\alpha\beta}^2$ term modifies the spin-2 propagator, while the $R^2$ term gives the correction to the spin-0 propagator.

We consider the same scattering as the Einstein gravity case, Eq.(\ref{the matter scattering}),
due to graviton exchanges in the $s$-, $t$-, and $u$-channels of Fig.\ref{fig:matter scattering}. 
The amplitude $M_{s} (\phi\phi\rightarrow\phi\phi)$ due to the $s$-channel exchange becomes
\begin{align}
M_{\rm s}=\frac {1}{\left(\beta s^2+\kappa^{-2}s\right)}\left(2tu- \frac{1}{3}(s-4m^2)^2 \right)-\frac{1}{3\left(2(3\alpha+\beta)s^2-\kappa^{-2} s\right)}(s+2m^2)^2.
\label{s-channel in R2}
\end{align}
In the Einstein gravity limit $\alpha=\beta=0$, this is reduced to Eq.(\ref{Ms}).
The $t$-channel exchange and $u$-channel exchange amplitudes are obtained from $M_{s}$ by the crossing relations
$s \leftrightarrow t$ and $s \leftrightarrow u$, respectively.
We add these two and the contact term to the $s$-channel exchange amplitude (\ref{s-channel in R2}) to obtain the amplitude $M(\phi\phi\rightarrow\phi\phi)$.

We examine the unitarity bound of this amplitude in the UV limit, i.e.,
\begin{align}
s\rightarrow\infty,~~{\rm with~\theta~being~fixed},
\end{align}
to compare it with that in the Einstein gravity. The scattering amplitude in the UV limit behaves as 
\begin{equation}
\begin{split}
M(\phi\phi\rightarrow\phi\phi)&\simeq-\lambda+\frac{2}{\beta}\left(\frac{tu}{s^{2}}+\frac{us}{t^{2}}+\frac{st}{u^{2}}\right)-\frac{3(2\alpha+\beta)}{2\beta(3\alpha+\beta)}\\
&=-\lambda+\frac{2}{\beta}g({\cos{\theta}})-\frac{3(2\alpha+\beta)}{2\beta(3\alpha+\beta)}\qquad\left(={\cal O}(s^{0})\right), 
\label{scattering amplitude in R2}
\end{split}
\end{equation}
where we have set $g(x)=\frac{1}{4}(1-x)(1+x)-2(1+x)/(1-x)^{2}-2(1-x)/(1+x)^{2}.$
It is gratifying to see that the amplitude (\ref{scattering amplitude in R2}) does not diverge, and hence,
unless the dimensionless parameter $\beta$ or $(3\alpha+\beta)$ is tuned small, it satisfies the unitarity bound at high energy,
in contrast with that in Einstein gravity. 

In deriving the asymptotic behavior (3.9) of $M$, $\beta \neq 0$ and $3\alpha + \beta \neq 0$ are implicitly 
assumed. The value of $M$ in the limit of $s\to \infty$ (as well as $t, u\to\infty$) for i) $\beta = 0$ and 
ii) $3\alpha + \beta = 0$ can be computed with some care. We give only the results:
\begin{align}
&\mbox{i)}\qquad M=2\kappa^{2}sf(\cos{\theta})+{\cal O}(s^{0})={\cal O}(s^{1}).\\
&\mbox{ii)}\qquad M=-\lambda+\frac{2}{\beta}f(\cos{\theta})-3+\frac{16}{3}\kappa^{2}m^{2}={\cal O}(s^{0}).
\label{3a+b}
\end{align}
Here $f(x)$ is defined below Eq.(\ref{s^1}). 
We note that the amplitude $M$ is divergent and hence the unitarity bound is not met for $\beta= 0$. 
This means that the unitarity bound is satisfied only if $(R_{\mu\nu})^2$  is included in the gravity action. 
Hence, in the title of this paper, we called it $(R_{\mu\nu})^2$  gravity rather than the usual $R^2$  gravity. 
Case ii) $3\alpha+\beta = 0$ is subtler. 
Though the UV behavior of the propagator is ${\cal O}(p^2 )$ and is not good in this case, 
the amplitude $M$ is finite (see Eq.(\ref{3a+b})) because the ${\cal O}(s^1 )$ terms in $M$ due to the $s$-, $t$-, $u$-channel exchanges cancel,
and hence the unitarity bound is met. 
\section{Discussion}
\label{sec:Discussion}
We have demonstrated the usefulness of matter scattering in the study of high-energy unitarity in gravity theory.
The scalar-field scattering amplitude $M(\phi\phi\rightarrow\phi\phi)$ in Einstein gravity is shown not to satisfy unitarity. 
We have shown that the matter scattering amplitude in $R^2$ gravity ($R_{\mu \nu}^{2}$ gravity, more precisely) satisfies the unitarity bound,
which is an inequality to be satisfied in any unitary theory.
This is due to the higher derivative term in the latter theory, and could be related to the renormalizability of this theory.

For the high-energy unitarity of scalar-field scattering, 
the $R_{\mu \nu}^{2}$ (or $C_{\mu\nu\alpha\beta}^2$) term is required. 
The $R^2$ term alone does not render the amplitude unitary. 
The $C_{\mu\nu\alpha\beta}^2$ and $R^2$ terms modify the high-energy behavior of the graviton propagators for 
spin-2 and spin-0, respectively. 
Hence, the modification of the spin-2 propagator is essential for the high-energy unitarity bound of scalar-field scattering to be satisfied. 
Even if the spin-0 part of the propagator behaves as ${\cal O}(p^2)$ in the UV limit, 
the total amplitude is suppressed by a constant. 
This happens since the ${\cal O}(p^2)$ parts of spin-0 exchanges in the $s$-, $t$-, $u$-channels cancel. 
This cancellation would be a generic property for the spin-0 exchange. 
The vertex for spin-0 does not have the spacetime index and thus the contraction of the spacetime index happens in each vertex. 
Therefore, the amplitude of  the $s$-channel, for instance, should be a function of only $s$. 
If the amplitude is ${\cal O}(p^2)$, the amplitude of the $s$-channel is $Cs$ in the UV limit, where $C$ is a constant.
The crossing relation sets the amplitudes of the $t$- and $u$-channels to be $Ct$ and $Cu$. 
Then, the sum of the three channels becomes $C(s+t+u)=4Cm^2$, i.e., it does not diverge in the UV limit.

In this letter we have considered matter scattering in Einstein and $R^2$ gravity. 
We have seen that, from the viewpoint of the scattering amplitudes for matter, the UV behavior improves.  
Investigation of the matter--graviton or graviton--graviton scattering would be interesting for the following two reasons.
Firstly, matter--graviton scattering in Einstein gravity is known not to obey unitarity \cite{BG}.
It is an urgent problem to see whether the matter--graviton scattering amplitude satisfies unitarity in $R^{2}$ gravity. 
Secondly, $R^{2}$ gravity has negative norm modes because of the higher derivative in the kinetic term for the metric components, and thus it is not unitary. 
The unitarity bound can be derived with the assumptions of unitarity, i.e. $SS^\dagger=1$ and no negative norm states.
Now, $R^{2}$ gravity leads to the well-behaved amplitude for matter scattering satisfying the unitarity bound, even with the negative norm states of gravitons. 
It would be interesting to see what happens for the scattering amplitudes involving gravitons in the external lines. 
(We are now analyzing a higher-order derivative scalar-field model that is an analog to $R^{2}$ gravity \cite{AIIKN}.)

We have shown that the UV behavior of the amplitude for $\phi$--$\phi$ scattering satisfies the unitarity bound due to modification of the UV behavior of the graviton propagator. 
There, it is essential that the graviton propagator behaves as $p^{-4}$ in the UV limit. 
This propagator with the higher-order derivative can be represented as 
\begin{eqnarray}
-\frac{1}{p^4-m^2 p^2} =\frac{1}{m^2} \left( \frac{1}{p^2}-\frac{1}{p^2-m^2} \right) . 
\end{eqnarray}
This corresponds to two propagators with opposite signs. 
Hence, a field with fourth-order derivatives can be equivalent to two fields with second-order derivatives, 
one of which creates negative norm states. 
The appearance of negative norm states is generic even if we introduce higher-order derivatives. 
Therefore, it seems to be inevitable that negative norm states appear in the theory with higher derivative terms.
The only possible way to avoid this problem of negative norm states would be to introduce the infinitely higher-order derivative terms, such as $\phi e^{-\Box} \phi$~\cite{TBM}. 
The structure of the poles in such theories is very different from that in the truncated approximation by the finite order of $-\Box$, and 
the propagator does not have additional poles due to the higher-order derivatives.
This would mean that quantum gravity theory has infinitely higher-order derivative terms, such as string field theory.
In string theory, scattering amplitudes are known to be finite, and the theory is unitary.
Whether the finiteness mechanism of the scattering amplitudes
in $R^{2}$ gravity has any similarity to that in string theory is an interesting question.
However, there are a few apparent obstacles.
String theory is constructed in $26$D spacetime, whereas our amplitude due to gravity interactions is in 4D spacetime.
Moreover, the complete form of string field theory including both closed and open strings is not known.
It would be interesting to compare our work with the recent developments in string field theory
 (see Ref. \cite{LEKSV}). 

It would be interesting to investigate how general the relation between renormalizability and high-energy unitarity is. 
One of the directions is the study of the relation to Lorentz symmetry. 
The relation in the case without Lorentz symmetry is confirmed \cite{FIIK1,FIIK2}.
The renormalizability of Ho$\check{\rm r}$ava-Lifshitz gravity~\cite{Horava} could be checked
by examining the unitarity of the non-relativistic scattering amplitudes \cite{Izumi}.
Computing matter--graviton scattering is possible and it will give a clue to the renormalizability question of Ho$\check{\rm r}$ava gravity.

\section*{Acknowledgments}
The authors would like to thank T. Noumi, Y.-T. Huang, and M. Horikoshi for valuable discussions.
T. I. thanks Yoonbai Kim for the kind hospitality during his six-month stay at Sungkunkwan University,
where he benefited greatly from the discussions with the theory people and the students.
K. I. is supported by a JSPS Grants-in-Aid for Young Scientists (B) (No.\,17K14281) and for Scientific Research (A)(No.\,17H01091).
T. K. is supported by a Waseda U. Grant-in-Aid for Special Research Projects (2017K-233).
The authors wish to thank T. Hatsuda of RIKEN, Y. Tsuboi of Chuo University, and C. S. Lim and H. Suzuki of Tokyo Woman's Christian University for their kind hospitality.
Y. A. wishes to thank S. Kawai of SKKU for support during his visit.

\end{document}